\RequirePackage{fix-cm}
\documentclass[twocolumn,epjc3]{svjour3}
\RequirePackage[T1]{fontenc}
\smartqed  
\RequirePackage{graphicx}
\RequirePackage{mathptmx}      
\RequirePackage{microtype}     
\RequirePackage{amsmath}
\RequirePackage{amssymb}
\RequirePackage{booktabs}
\RequirePackage{multirow}
\RequirePackage{flushend}      
\RequirePackage[numbers,sort&compress]{natbib}
\RequirePackage[colorlinks,citecolor=blue,urlcolor=blue,linkcolor=blue]{hyperref}

\makeatletter
\def\epjc@institutelabel#1{\safelabel{#1thanks}\thanks@hypertarget{#1}}
\g@addto@macro\institute@start{\let\label\epjc@institutelabel}
\makeatother

\newcommand*{\doi}[1]{DOI \href{https://doi.org/#1}{{\urlstyle{rm}\nolinkurl{#1}}}}

\journalname{Eur. Phys. J. C}

\begin{document}

\title{Extrapolation and explainability in hadron mass predictions with machine learning}


\author{Esra Y\"uksel\thanksref{e1,addr1}
        \and
        H\"useyin Bahtiyar\thanksref{e2,addr2}
        \and
        Derya Soydaner\thanksref{e3,addr3}
}

\thankstext{e1}{e-mail: e.yuksel@surrey.ac.uk}
\thankstext{e2}{e-mail: huseyinbahtiyar@gmail.com}
\thankstext{e3}{e-mail: d.soydaner@liacs.leidenuniv.nl}

\institute{School of Mathematics and Physics, University of Surrey, Guildford, Surrey GU2 7XH, United Kingdom\label{addr1}
           \and
           Independent Researcher, Istanbul, T\"urkiye\label{addr2}
           \and
           Leiden Institute of Advanced Computer Science, University of Leiden, Leiden, The Netherlands\label{addr3}
}

\date{Received: date / Accepted: date}

\maketitle

\begin{abstract}
In this work, we study the predictive performance of four machine learning (ML) models for hadron masses, with particular attention to extrapolation beyond the conventional hadrons represented in the training data. Calculations are performed using a multilayer perceptron (MLP), eXtreme Gradient Boosting (XGBoost), Light Gradient Boosting Machine (LightGBM), and Gaussian process regression (GPR). ML models are trained on conventional hadrons, using features that describe valence quark content and quantum numbers. Then, we evaluate their performance on test sets and on a separate extrapolation set of 39 states, including exotic candidates and heavy baryons, none of which appear in the training or test sets. We find that the MLP achieves the lowest extrapolation Root Mean Square Error (RMSE) on the $30$ states with measured masses at  $294.2 \pm 57.9$~MeV, roughly one-third that of the other models. We also show that the MLP reproduces the overall mass scale of nine unobserved doubly heavy baryons predicted by theoretical calculations, although it underestimates the doubly bottom masses. We use SHapley Additive exPlanations (SHAP) analysis and show that heavy quark content is the most significant factor in the MLP's mass predictions, with the remaining features contributing differently across hadron types. Our results quantify the extrapolation accuracy of ML models for hadron masses and identify the features that drive the MLP predictions.
\end{abstract}

\section{Introduction}
Mass is a fundamental observable that carries important information about hadrons' nature and structure, as well as the dynamics of the strong interaction. Accurate measurements and calculations of hadron masses therefore provide an important test of our understanding of the strong interaction and its dynamics, and a benchmark for theoretical calculations. To date, phenomenological models \cite{PhysRevD.32.189, PhysRevD.34.2809} and lattice Quantum Chromodynamics (QCD) calculations \cite{doi:10.1126/science.1163233,PhysRevD.84.074508} have achieved considerable success in describing mesons and baryons. However, the results of these theoretical approaches differ when one goes beyond ordinary mesons and baryons, i.e. for exotic hadrons. Furthermore, lattice QCD calculations are computationally demanding, and additional difficulties arise for excited states and resonances close to strong-decay thresholds \cite{LUSCHER1991531, PhysRevLett.111.192001}.

In recent years, advances in experimental facilities have enabled us to observe many exotic hadron candidates as well as conventional hadrons containing multiple heavy quarks. The Belle Collaboration performed an important experiment in 2003 that resulted in the discovery of the charmonium-like state $X(3872)$ \cite{Belle:2003nnu}. The discovery of $X(3872)$ has also motivated other experimental and theoretical studies of unconventional $XYZ$ hadron states \cite{BRAMBILLA20201, RevModPhys.90.015003}. In 2013, the LHCb Collaboration also determined the quantum numbers of $X(3872)$ as $J^{PC}=1^{++}$ \cite{LHCb:2013kgk}. Theoretically, different approaches have been used to understand the internal structure of the $X(3872)$ better, including a compact tetraquark description \cite{PhysRevD.71.014028}, a hadronic-molecule interpretation \cite{PhysRevD.69.074005}, or a mixed charmonium--molecule description \cite{PhysRevD.80.056002}; see also Refs. \cite{RevModPhys.90.015004,LEBED2017143}. However, its internal structure remains inconclusive.

Following the discovery of $X(3872)$, other exotic hadron candidates, such as $Z_{c}$ \cite{BESIII:2013ris}, $Z_{b}$ \cite{Belle:2011aa}, and hidden-charm pentaquarks \cite{LHCb:2015yax,LHCb:2019kea}, have also been reported. The observation of the doubly charmed baryon $\Xi_{cc}^{++}$ by the LHCb Collaboration in 2017 \cite{LHCb:2017iph} provided an important experimental benchmark for earlier quark-model and lattice-QCD predictions of doubly heavy baryons \cite{PhysRevD.90.094007, PhysRevD.90.094507}. Recently, the doubly charmed tetraquark $T_{cc}^{+}$ has been observed by LHCb and has provided another important example of an exotic heavy flavour hadron \cite{Aaij2022}. Meanwhile, the observation of additional excited heavy baryon states, including new $\Omega_b^-$ \cite{LHCb:2020tqd}, $\Omega_c^0$ \cite{LHCb:2023sxp}, and $\Xi_b$ states \cite{LHCb:2023zpu} has further broadened the experimentally verified hadron spectrum.

In addition to progress in theoretical and experimental research, machine learning (ML) has also been increasingly applied across many areas of physics over the past decade due to the rapid growth of available data, advances in computing hardware, and progress in ML algorithms. ML models provide a flexible framework for predicting hadronic properties and can learn complex relationships in experimental or theoretical data. They have demonstrated promising performance in nuclear physics \cite{GAZULA19921, ATHANASSOPOULOS2004222, PhysRevLett.124.162502, Yuksel:2021nae, Jiao:2018dro, Yuksel:2024zky}, particle physics \cite{Guest:2018yhq, Radovic:2018dip, Aaij:2012me, Luo:2017ncs, NEXT:2016ire, Psihas:2020pby, Dudko:2020qas, Sombillo:2021ifs, Bahtiyar:2022une, Rostami:2025sff, Malekhosseini:2024eot, Rostami:2026jyz}, and lattice QCD \cite{Shanahan:2018vcv, Yoon:2018krb, Zhou:2023pti}. Although ML models are powerful tools for making predictions on various data, the scientific community also criticizes them for operating as black boxes and for making it difficult to understand how they learn physical information for their predictions. In hadron spectroscopy, making accurate mass predictions alone is not sufficient. We also need to assess which input features have the greatest influence on those predictions. Recent advances in explainable artificial intelligence (XAI) also provide tools to address this issue by quantifying each feature's contribution to model predictions \cite{Lundberg:2017, shapley1953value}.

In this study, we evaluate whether ML models can infer mass relations using only standard mesons and baryons, and then apply these relations to predict the masses of exotic candidates and heavy baryons that were left out of the training set. To this end, we compare the predictive performance of four ML models: MLP, GPR, XGBoost, and LightGBM. Each model is trained on the measured masses of $432$ conventional hadron entries using the same $15$ physically motivated features. We evaluate their performance over ten common random train--test partitions and on a separate set of $39$ states excluded from both training and testing. The extrapolation set consists of $30$ states with measured masses and nine unobserved doubly heavy baryons. We use these states to examine predictions for quark configurations absent from the training data. We also use SHAP analysis to examine the feature contributions to the MLP predictions.

This paper is organized as follows. Section~\ref{sec:impandmodel} describes the experimental dataset, feature selection, and machine learning models. Section~\ref{sec:res} presents the prediction results and the SHAP-based interpretation. Finally, Sect.~\ref{sec:discuss} summarizes the conclusions.

\section{Methodology}
\label{sec:impandmodel}

\subsection{Dataset}
\label{subsec:dataset}

We obtain experimentally measured hadron masses from the 2020 Particle Data Group (PDG) compilation~\cite{ParticleDataGroup:2020ssz} using the Particle Python package~\cite{rodrigues2020scikit}. Then, we update selected masses with values from the 2026 PDG compilation~\cite{ParticleDataGroup:2026mpi}. The dataset consists of the masses of $432$ conventional mesons and baryons. We exclude exotic hadron candidates and doubly heavy baryons from the dataset and use them exclusively as an independent out-of-sample prediction set to assess the extrapolation performance of ML models. We also use hadron states listed in the 2026 PDG compilation but absent from the training dataset as additional out-of-sample benchmarks. Due to the small size of the experimental dataset, we perform a random split and allocate 80\% for training and 20\% for testing. This allows us to assess performance on unseen data while keeping most examples available for training.

For each hadron, our dataset contains 15 physics-based input features: parity $P$, total angular momentum $J$, isospin $I$, electric charge $Q$, as well as ten variables describing the valence quark content, and the state index $n_{\rm state}$. We use hadron mass as the target for ML models. To describe the quark content of hadrons, we use ten integer-valued variables giving the numbers of $u$, $d$, $s$, $c$, and $b$ quarks and their respective antiquarks, which in turn give the total number of quarks or antiquarks in the hadron. The parity $P$ of hadrons is taken from the adopted $J^P$ assignment of each hadronic state and $P=\pm1$. In cases where $J^P$ has not been experimentally established, the assignment follows the PDG convention or a theoretically motivated value. The isospin $I$ and total angular momentum $J$ take non-negative integer or half-integer values depending on the hadronic state. The electric charge $Q$ is expressed in units of the elementary charge $e$ and calculated using the valence quark content. In the present dataset, $Q$ takes the integer values $-2, -1, 0, +1,$ and $+2$. Finally, the feature vector is written as

\begin{equation}
\mathbf{x}_{\rm base}
=
\left(
n_u,n_d,n_s,n_c,n_b,
n_{\bar u},n_{\bar d},n_{\bar s},n_{\bar c},n_{\bar b},
P,J,I,Q,n_{\rm state}
\right).
\end{equation}

It should be noted that hadrons can have identical quark content and quantum numbers $(I,J,P,Q)$ but different masses. To distinguish such states, we introduce an additional state index $n_{\rm state}$ \cite{Gal:2022yqu}. For hadrons with identical quark content and quantum numbers, the states are ordered according to their masses. The lightest state is assigned $n_{\rm state}=0$, the next lightest $n_{\rm state}=1$, and so on. For instance, the proton and the $N(1440)^{+}$ both have quark content $uud$ and quantum numbers $I=1/2$, $J=1/2$, $P=+1$, and $Q=+1$. Their feature vectors are

\begin{equation}
\mathbf{x}_{p}
=
\left(
2,1,0,0,0,\,
0,0,0,0,0,\,
1,\frac{1}{2},\frac{1}{2},1,0
\right),
\end{equation}

and

\begin{equation}
\mathbf{x}_{N(1440)^{+}}
=
\left(
2,1,0,0,0,\,
0,0,0,0,0,\,
1,\frac{1}{2},\frac{1}{2},1,1
\right).
\end{equation}

Without the state index, these two states would have identical input features despite having different experimentally measured masses.

\subsection{Machine Learning Models}
\label{subsec:ML}

In this study, we compare four machine learning models: MLP, XGBoost, LightGBM, and GPR. All four models use the same 15-feature input representation and predict a single hadron mass as output. Since hadron mass is a continuous quantity, predicting it is a regression task. The models may differ in their predictions for hadrons within and beyond the region represented by the training data. We examine these differences through the extrapolation test in Sect.~\ref{sec:extrap}. The hyperparameters of all models are selected using five-fold cross-validation within the reference training subset, with identical shuffled folds generated using a random seed of 42. Performance is summarized as the mean and sample standard deviation over ten random 80\%--20\% train--test partitions, using the same partitions for all models and keeping the hyperparameters fixed.

\paragraph{Multilayer perceptron (MLP).} An MLP is a feedforward neural network that is well suited to our regression problem. The parameters of an MLP are learned during training by minimizing a loss function. This function measures the difference between the target values and the model predictions. Starting from the network architecture of our previous study~\cite{Bahtiyar:2022une}, we optimize the hyperparameters using five-fold cross-validation on the training subset. Unlike Ref.~\cite{Bahtiyar:2022une}, we train without data augmentation and use a smooth $L_1$ loss. To reduce the risk of overfitting, the number of model parameters is kept below the number of training samples~\cite{goodfellow2016}, following the criterion discussed in our previous work ~\cite{Bahtiyar:2022une}. As shown in Table~\ref{tab:hp_15f}, our MLP has one hidden layer with 16 units, takes 15 features as input, and predicts a single hadron mass as the output. 

Training an MLP involves two main steps. First, during the forward pass, the input features are multiplied by weights and summed, and the resulting values are passed through an activation function in the hidden layer. In our MLP, we use $\tanh$ as the activation function. The output of the hidden layer is then passed to the output layer, where a linear function is used to predict the hadron mass. In this way, the MLP learns a continuous nonlinear relation between the input features and the predicted hadron mass.  

The prediction error is then calculated using a loss function. We use the smooth $L_1$ loss, which behaves quadratically for small errors and linearly for large ones. Therefore, it is less sensitive to large errors than the commonly used mean squared error. Second, during the backward pass, the gradients of the loss with respect to the parameters are calculated. These gradients are then used to update the parameters with the aim of reducing the prediction error. For these updates, we use the Nadam algorithm~\cite{dozat2016incorporating} as the optimizer. Nadam (Nesterov-accelerated adaptive moment estimation) combines the Adam optimizer~\cite{kingma2014adam} with Nesterov momentum. The training data are shuffled and divided into minibatches of 32 samples at each epoch. Further details on optimizers and adaptive learning algorithms can be found in~\cite{soydaner2020comparison}, while the architecture and training of MLPs are discussed in~\cite{bahtiyar2022application}. Finally, the fitted function can also be evaluated for previously unseen hadron configurations, allowing us to assess its ability to extrapolate to novel quark contents.

\begin{table}[ht]
\centering
\small
\caption{Model configurations for ML models.
Hyperparameters for all four models were selected by five-fold
cross-validation within the 345 training hadrons, with the 87 test hadrons evaluated once at the end.}
\label{tab:hp_15f}
\begin{tabular}{llr}
\toprule
Model & Parameter & Value \\
\midrule
\multirow{9}{*}{MLP}
 & architecture          & $15\to16\to1$ \\
 & activation            & $\tanh$ \\
 & initialization        & Glorot unif. \\
 & optimizer             & Nadam \\
 & learning rate         & $10^{-3}$ \\
 & weight decay          & $10^{-4}$ \\
 & batch size            & 32 \\
 & loss                  & smooth $L_1$ \\
 & epochs, max (sel.)    & 4500 (2221) \\
\midrule
\multirow{7}{*}{GPR}
 & kernel        & $C\times$Mat\'ern ARD \\
 & smoothness    & $\nu=1/2$ \\
 & $\alpha_0$    & 0.05 \\
 & length-scale bounds & $[0.05,\,10^{4}]$ \\
 & amplitude bounds    & $[10^{-3},\,10^{3}]$ \\
 & density weighting   & on, $k=5$ \\
 & ensemble size $M$, restarts & 3, 8 \\
 \midrule
 \multirow{10}{*}{XGBoost}
 & n\_estimators      & 800 \\
 & max\_depth        & 3 \\
 & learning\_rate     & 0.10 \\
 & subsample          & 1.00 \\
 & colsample\_bytree  & 1.00 \\
 & min\_child\_weight & 15 \\
 & reg\_alpha         & 0.01 \\
 & reg\_lambda       & 5.0 \\
 & gamma            & 0 \\
 & max\_delta\_step   & 5 \\
\midrule
\multirow{11}{*}{LightGBM}
 &  n\_estimators       & 2600 \\
 &  max\_depth          & $-1$ (unlimited) \\
 &  num\_leaves         & 7 \\
 &  learning\_rate     & 0.02 \\
 &  subsample          & 1.00 \\
 &  subsample\_freq     & 0 (no bagging) \\
 &  colsample\_bytree   & 0.80 \\
 & min\_child\_samples & 10 \\
 & reg\_alpha          & 0 \\
 & reg\_lambda        & 0 \\
 & min\_split\_gain    & 0 \\
\bottomrule
\end{tabular}
\end{table}

\paragraph{Gradient-boosted decision trees.}

In addition to the MLP, we use tree-based ML algorithms for hadron mass prediction. These algorithms are built on \textit{decision trees}, which are hierarchical models composed of internal decision nodes and terminal nodes, called \textit{leaves}. Given an input, a test is applied at each decision node, and one of the branches is followed depending on the outcome. This process starts at the root and is repeated recursively until a leaf is reached, whose value provides the final prediction~\cite{alpaydin2021machine}. Decision trees can be used for both classification and regression, with the tree constructed by recursively partitioning the data according to the input features. The prediction of a regression tree is therefore piecewise constant and takes one of its fitted leaf values. This matters for extrapolation: in the training sample, no hadron contains more than one charm or one bottom quark, so every split on $n_c$ or $n_b$ lies at a threshold below two, and a doubly heavy state is routed to the same leaf as its singly heavy counterpart. 

Gradient boosting is an ensemble learning method that combines multiple decision trees to improve prediction performance. It uses the principle of \textit{boosting}~\cite{schapire1990strength}, in which predictors are trained sequentially, with each subsequent model aiming to reduce the residual errors of the ensemble~\cite{geron2022hands}. In \textit{gradient boosting}, new trees are added incrementally to the ensemble to minimize a loss function. In this work, we use two gradient-boosted decision tree algorithms: XGBoost and LightGBM.

eXtreme Gradient Boosting (XGBoost)~\cite{chen2016xgboost} constructs decision trees sequentially. XGBoost includes both $L_1$ and $L_2$ regularization, controlled by the $reg\_alpha$ and $reg\_lambda$ hyperparameters, respectively, as shown in Table~\ref{tab:hp_15f}. The number of estimators corresponds to the number of boosting rounds, or decision trees added to the ensemble. In our model, we use 800 trees, each with a maximum depth of 3, which limits the complexity of the individual trees. The learning rate is set to 0.10 and controls the contribution of each new tree to the ensemble.

Light Gradient Boosting Machine (LightGBM)~\cite{ke2017lightgbm} uses a \textit{leaf-wise} growth strategy, i.e. the leaf that provides the largest reduction in the loss is selected for splitting. In our model, the maximum number of leaves is set to 7, while no explicit maximum tree depth is specified. We use 2600 estimators with a learning rate of 0.02. A minimum of 10 data points is required in each leaf. Unlike our XGBoost configuration, which uses all input features, the LightGBM model randomly selects 80\% of the input features for constructing each tree. In our implementations, the two models differ mainly in how trees are grown: XGBoost uses a level-wise strategy, while LightGBM uses a leaf-wise strategy.
  
\paragraph{Gaussian process regression (GPR).} GPR is a probabilistic ML method that models the relation between input features and target values using a Gaussian process~\cite{mackay1998introduction,rasmussen2006gaussian}. It starts with a prior distribution over functions, characterized by a mean function and a kernel function, with Gaussian-distributed function values at input points. We use Bayes' theorem to update the prior distribution with the training data and obtain the posterior distribution. This posterior allows GPR to make predictions for unseen inputs and also estimate their associated uncertainties. 

In our implementation, we model $\ln(m/\mathrm{GeV})$ using a constant kernel multiplied by a Mat\'ern kernel with $\nu=1/2$ and a separate length scale for each feature. The input features and target values are standardized using statistics from the training set only. Each training hadron $i$ is assigned a noise variance $\alpha_i=\alpha_0/w_i$ in standardized log-mass units, where $w_i$ is its mean distance to its $k=5$ nearest neighbors in the standardized feature space, normalized to unit mean and clipped to $[0.1,10]$, so that hadrons in sparsely populated regions are fitted more closely. The final hyperparameters and bounds are listed in Table~\ref{tab:hp_15f}. For the final predictions, we use an ensemble of three GPR models with the same configuration but different randomly drawn initial kernel hyperparameters and optimizer seeds. Each model is fitted to each training set by maximizing the log marginal likelihood with respect to the kernel amplitude and length scales, using eight optimizer restarts. The predicted hadron mass is $\widehat m=\exp(\mu)\,\mathrm{GeV}$, where $\mu$ is the average of the three predicted log-masses, corresponding to the median of the log-normal predictive distribution.


\subsection{Explainable AI} 
ML models can provide accurate predictions. However, understanding how the input features influence these predictions is often difficult. For this reason, ML models are often regarded as `black boxes'. Explainable artificial intelligence (XAI) methods aim to address this issue by providing insights into the behavior of trained ML models. Among the various XAI approaches, we use SHapley Additive exPlanations (SHAP)~\cite{Lundberg:2017}, a widely used technique based on Shapley values from cooperative game theory~\cite{shapley1953value}. 

In this technique, the input features are treated as `players' that contribute to the model prediction. SHAP assigns a contribution to each input feature for a given prediction, indicating how that feature shifts the predicted value relative to a reference prediction. By examining these contributions across hadrons, we can identify which features have the greatest influence on the predicted masses and how their effects vary among different hadron configurations. This allows us to investigate how the ML models use the properties of hadrons when making their predictions. 

In this study, we apply SHAP to the model that achieves the best extrapolation performance among the four ML models. We calculate exact interventional Shapley values for the MLP trained on the reference split, and using a background set of 30 training hadrons. To examine global feature importance, we calculate the mean absolute SHAP value of each input feature across the 432 conventional hadrons. We also examine local SHAP values for selected exotic candidates and doubly heavy baryons to understand how individual features contribute to their predicted masses.

\section{Results and Discussion}
\label{sec:res}

\subsection{Predictive performance of the ML models}
\label{subsec:pred}

We first evaluate the predictive performance of the four ML models on the training and test sets of conventional hadrons. The random train-test split primarily assesses how well the models interpolate within the conventional-hadron dataset. It does not directly assess the extrapolation performance of ML models to hadrons outside the training distribution, such as exotic candidates or doubly heavy baryons.

In Table~\ref{tab:test_performance}, we summarize the training- and test-set root mean squared errors (RMSEs) obtained using four ML models and 15 features. We find that GPR achieves the lowest mean training and test RMSEs, at $90.3 \pm 3.6$~MeV and $169.3 \pm 45.2$~MeV, respectively. However, its mean test RMSE is also very close to those of XGBoost and LightGBM, and it shows the largest variation across runs. MLP has the highest mean test RMSE but is also close to other models. These results show that all four models capture a useful relationship between the 15 physically motivated features and hadron masses within the interpolation regime.

\begin{table}[htbp!]
\centering
\caption{Training- and test-set RMSE values of the four ML models using 15 input features. Results are reported as the mean
$\pm$ standard deviation over 10 runs with different random seeds. All values are expressed in units of MeV.}
\label{tab:test_performance}
\begin{tabular}{@{}lcc@{}}
\toprule
Model & Training RMSE & Test RMSE\\
\midrule
MLP      & $151.0 \pm 7.8$ & $198.0 \pm 19.6$ \\
GPR      & $ 90.3 \pm 3.6$ & $169.3 \pm 45.2$ \\
XGBoost  & $121.5 \pm 4.0$ & $175.2 \pm 12.3$ \\
LightGBM & $101.0 \pm 4.1$ & $176.4 \pm 17.1$ \\
\bottomrule
\end{tabular}
\end{table}

To compare model performance across hadron sectors, we calculate the RMSE separately for mesons and for baryons. We use the same training and test split for all models within each run to ensure a fair comparison. The results are given in Table~\ref{tab:meson_baryon_rmse}.
All four models predict lower mean test RMSEs for baryons than for mesons in this dataset. LightGBM, XGBoost, and GPR give similar mean RMSEs on the meson test set, while GPR achieves the lowest mean test RMSE for baryons. However, we cannot claim a universal predictive advantage of models for the baryon sector. Unequal numbers of mesons and baryons in our dataset, together with differences in spectral composition, may also contribute to the observed differences in the predictive performance of models.

\begin{table}[th!]
\centering
\setlength{\tabcolsep}{3.5pt}
\caption{RMSE values for the meson and baryon subsets of the training and test data using the 15 feature representation. Results are reported as the mean
$\pm$ standard deviation over 10 random 
train-test splits.  All values are expressed in units of MeV.}
\label{tab:meson_baryon_rmse}
\begin{tabular}{@{}lcccc@{}}
\toprule
& \multicolumn{2}{c}{Training RMSE}
& \multicolumn{2}{c}{Test RMSE} \\
\cmidrule(lr){2-3}\cmidrule(lr){4-5}
Model
& Mesons & Baryons
& Mesons & Baryons \\
\midrule
MLP      & $165.8 \pm 14.5$ & $142.0 \pm 5.9$
         & $201.1 \pm 26.7$ & $194.3 \pm 31.2$ \\
GPR      & $41.6 \pm 2.9$ & $108.2 \pm 4.9$
         & $189.0 \pm 97.6$ & $149.0 \pm 16.8$ \\
XGBoost  & $130.3 \pm 7.9$ & $116.3 \pm 3.8$
         & $187.5 \pm 19.9$ & $167.5 \pm 17.2$ \\
LightGBM & $102.3 \pm 7.3$ & $100.2 \pm 4.5$
         & $183.4 \pm 35.6$ & $171.2 \pm 13.8$ \\
\bottomrule
\end{tabular}
\end{table}

\subsection{Extrapolation performance of the ML models}
\label{sec:extrap}

After assessing the performance of the four ML models on the training and test data, we now assess their predictions for 39 hadrons that were withheld from the training process. This set consists of 17 exotic candidates, 13 measured heavy baryons and 9 unobserved heavy baryons. These states allow us to assess the generalization performance of ML models beyond the training data, including extrapolation to constituent configurations absent from the training set.

\begin{table}[th!]
\centering
\caption{Extrapolation RMSEs for the 30 states with measured masses using ML models used in this study. The results are given for the full set and separately for 17 exotic candidates and 13 heavy baryons. Values are in MeV and represent the mean $\pm$ sample standard deviation over ten random train--test splits.}
\label{tab:extrap_class}
\begin{tabular}{@{}lccc@{}}
\toprule
Model & All (30) & Exotic (17) & Heavy baryons (13) \\
\midrule
MLP      & $294.2 \pm 57.9$ & $283.1 \pm 66.8$   & $304.3 \pm 66.5$ \\
GPR      & $780.9 \pm 23.5$ & $848.9 \pm 21.2$   & $681.4 \pm 33.4$ \\
XGBoost  & $844.1 \pm 32.7$ & $976.8 \pm 43.8$   & $629.4 \pm 17.0$ \\
LightGBM & $877.7 \pm 82.0$ & $1020.5 \pm 121.7$ & $642.0 \pm 22.4$ \\
\bottomrule
\end{tabular}
\end{table}

Exotic hadron candidates are known as states whose internal structure may differ from the conventional quark--antiquark or three-quark configurations. In the multiquark assignments considered here, the $X(3872)$ and $T_{cc}(3875)^+$ are tetraquark candidates with two quarks and two antiquarks. The $P_c(4312)^+$ is also a pentaquark candidate with four quarks and one antiquark. In this study, the ML models are only trained on conventional mesons and baryons with valence configurations $q\bar q$ and $qqq$, respectively. Therefore, ML predictions for these exotic hadrons test whether the mass relationships learned from conventional hadrons extend to systems with more valence constituents. This provides us with a direct assessment of the extrapolation abilities of ML models to constituent configurations absent from the training set.

\begin{figure*}[t!]
    \centering
    \includegraphics[width=0.84\linewidth]{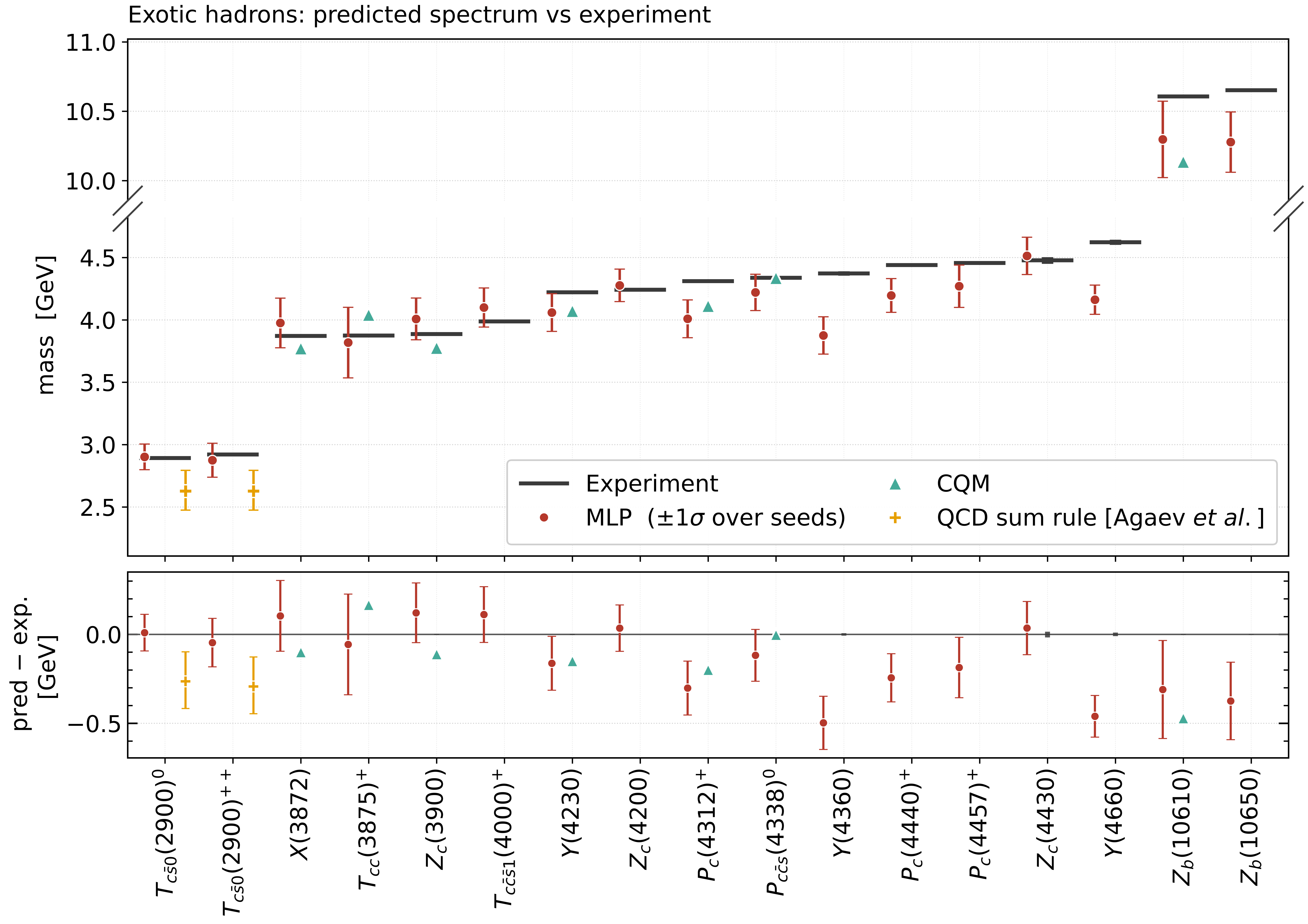}
\caption{Mass predictions for 17 exotic hadron candidates. Red circles show MLP predictions averaged over ten random train-test splits. Black horizontal lines indicate experimental masses from Ref.~\cite{ParticleDataGroup:2026mpi}. Green triangles show quark-model and coupled-channel predictions: the effective-quark-mass constituent quark model tabulated in Ref.~\cite{Gal:2022yqu} for $X(3872)$ and the $Z_c$, $Y$, $Z_b$, and $P_c$ states; the relativized quark model of Ref.~\cite{Lu:2020rog} for $T_{cc}(3875)^+$; and the coupled-channel calculation of Ref.~\cite{PhysRevD.108.074012} for $P_{c\bar{c}s}(4338)^0$.  Orange crosses show the QCD sum-rule result of Ref.~\cite{Agaev2018} for $T_{c\bar{s}0}(2900)$. The lower panel shows $m_\mathrm{pred}-m_\mathrm{exp}$.}
\label{fig:spectrum_exotic} 
\end{figure*}

Heavy baryons consist of three quarks, with at least one charm or bottom quark. In this study, we consider 22 heavy baryons, of which 10 are singly heavy excited states ($\Omega_c$, $\Xi_b$, and $\Omega_b$), and 12 are doubly heavy baryons containing two charm or two bottom quarks. The singly heavy states have quark contents that appear in the training set but lie well above the corresponding ground states. Therefore, they test extrapolation to higher excitations of known quark configurations. In our study, none of the training baryons contains two heavy quarks; hence, the doubly heavy subset tests extrapolation to new heavy quark configurations. Three of these, $\Xi_{cc}^{+}$, $\Xi_{cc}^{++}$ and $\Omega_{cc}^{+}$, have measured masses, and the remaining nine have not yet been observed. The exotic candidates test different numbers of valence constituents, the doubly heavy baryons test different heavy-quark content, and the singly heavy states test higher excitation energies. Reference masses for the exotic candidates and the measured heavy baryons are taken from the 2026 Review of Particle Physics~\cite{ParticleDataGroup:2026mpi}. The data for $\Xi_{cc}^{+}$ and $\Omega_{cc}^{+}$ are taken from ~\cite{LHCb_Xicc_plus_2026, LHCb_Omegacc_plus_2026} and these states were observed by LHCb after the PDG 2026 compilation.

For each ML model, we calculate the RMSE over the out-of-sample states for each of ten random train--test splits and report the mean and standard deviation of these ten values in Table~\ref{tab:extrap_class}. 

\begin{figure*}[t!]
    \centering
    \includegraphics[width=0.84\linewidth]{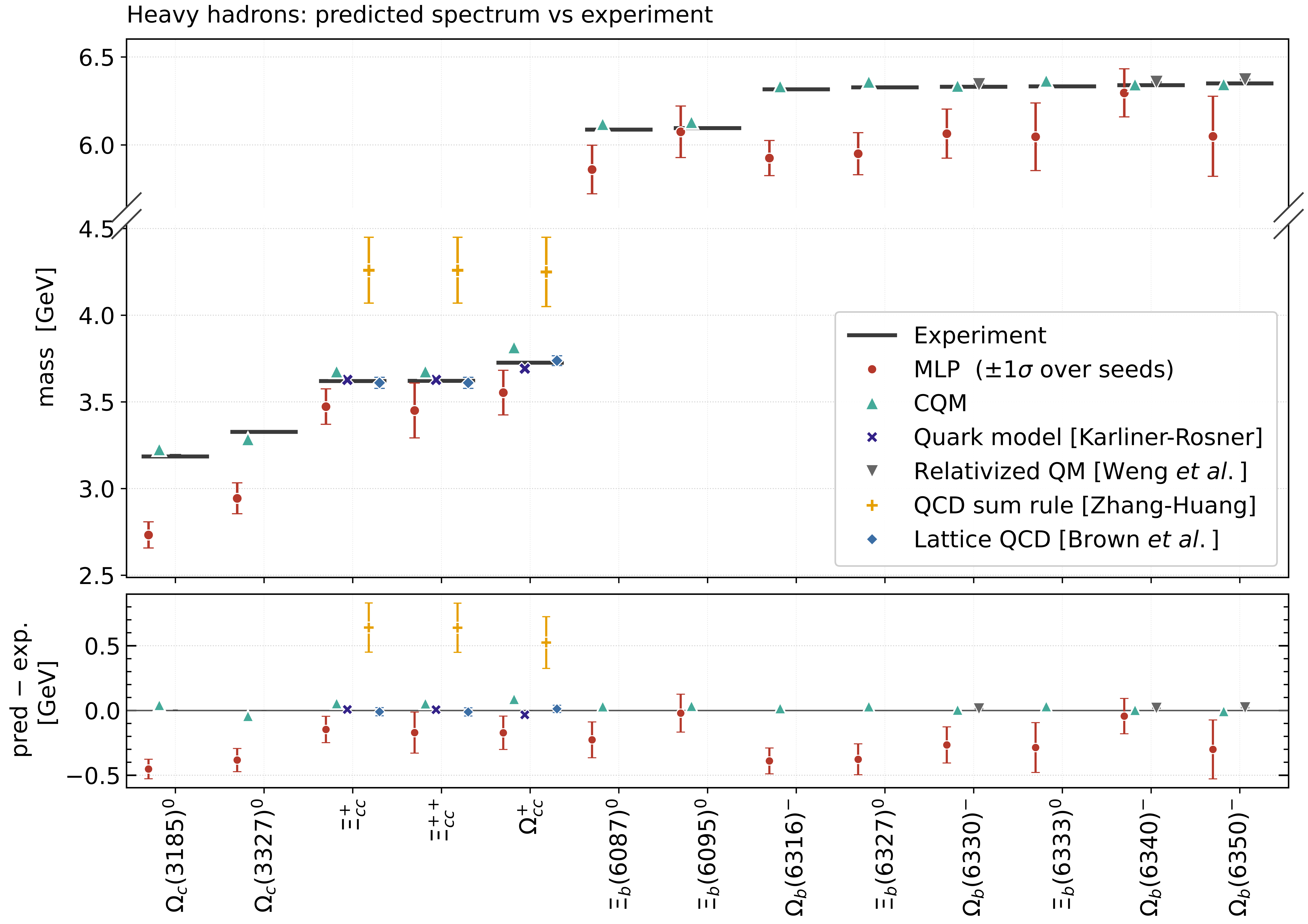}
\caption{Same as Fig.~\ref{fig:spectrum_exotic}, but for 13 heavy baryons. Black horizontal lines show experimental masses from the PDG~\cite{ParticleDataGroup:2026mpi}, supplemented by the LHCb measurements of $\Xi_{cc}^+$~\cite{LHCb_Xicc_plus_2026} and $\Omega_{cc}^+$~\cite{LHCb_Omegacc_plus_2026}. Green triangles show CQM predictions from Refs.~\cite{Roberts:2007ni,Yoshida:2015tia,Ebert:2007nw}; purple crosses show Karliner--Rosner quark-model predictions~\cite{PhysRevD.90.094007,PhysRevD.97.094006}; and grey inverted triangles show predictions from the relativized quark model of Weng \emph{et al.}~\cite{PhysRevD.110.056052}. Orange crosses show QCD sum-rule predictions by Zhang and Huang~\cite{PhysRevD.78.094007}. Blue diamonds show lattice-QCD results from Brown \emph{et al.}~\cite{PhysRevD.90.094507}.}
\label{fig:spectrum_heavy}
\end{figure*}

\begin{figure*}[t!]
    \centering
\includegraphics[width=0.82\linewidth]{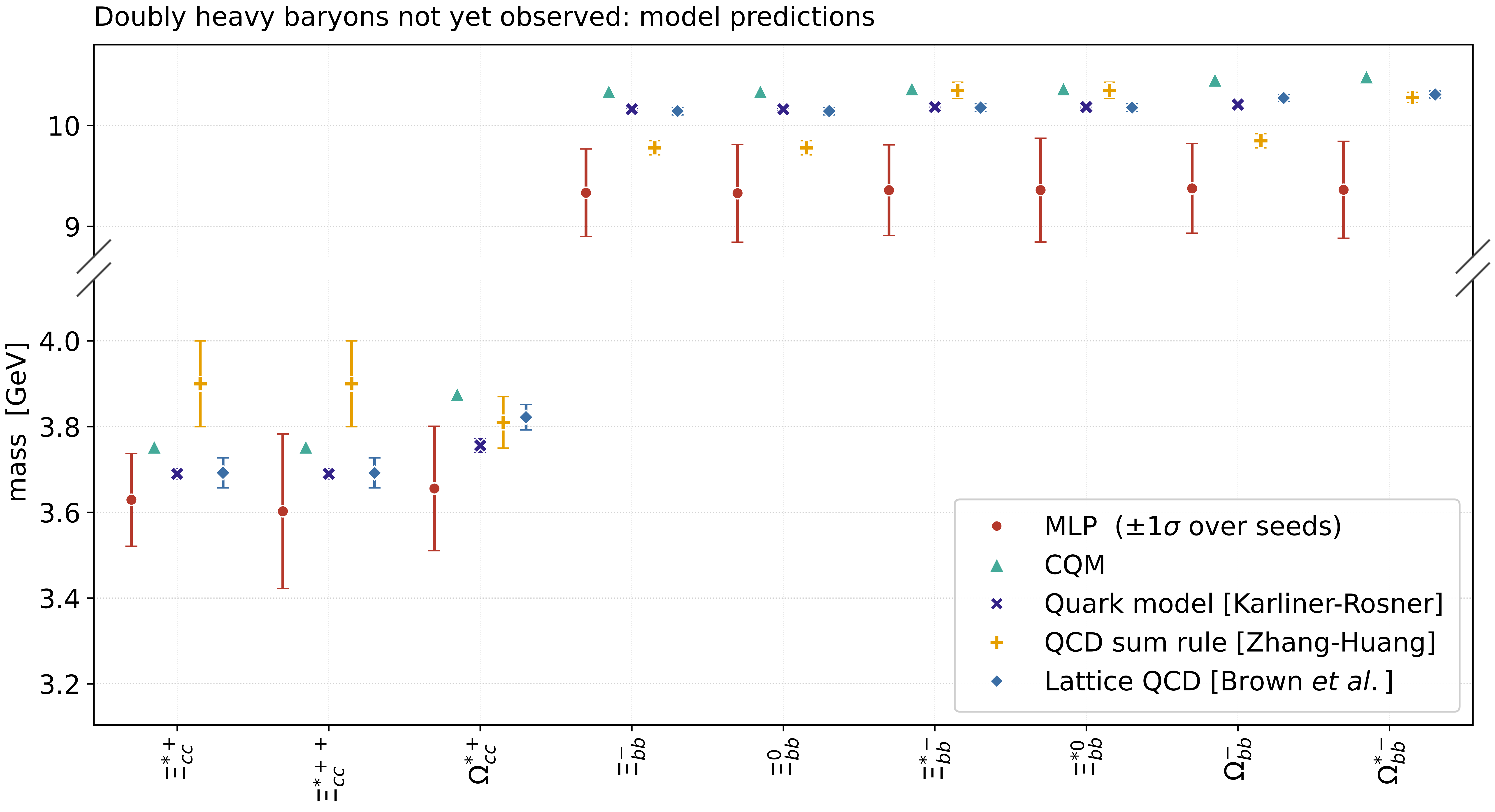}
\caption{Mass predictions for nine doubly heavy baryons without experimental reference masses. Red circles show MLP predictions averaged over ten random train-test splits. Green triangles show constituent-quark-model predictions by Roberts and Pervin~\cite{Roberts:2007ni}; purple crosses show Karliner--Rosner quark-model predictions~\cite{PhysRevD.90.094007,PhysRevD.97.094006}; orange crosses show QCD sum-rule predictions by Zhang and Huang~\cite{PhysRevD.78.094007}; and blue diamonds show lattice-QCD results by Brown \emph{et al.}~\cite{PhysRevD.90.094507}.}
\label{fig:spectrum_nonobserved}
\end{figure*}

We find that the MLP gives the best overall results with the lowest mean RMSE of $294.2 \pm 57.9$ MeV for all 30 states excluded from the training set. The other three models are less accurate in their predictions and mean RMSEs approximately 2.7-3.0 times that of MLP. In Table~\ref{tab:extrap_class}, we also present differences in variation across runs. GPR, XGBoost, and LightGBM have standard deviations of approximately $3\%$, $4\%$, and $9\%$ of their overall mean RMSEs, respectively. These models produce more stable errors across splits; however, they predict masses less accurately than the MLP. We find that the MLP has a larger relative standard deviation of approximately $20\%$ and this variation may reflect sensitivity to both the training partition and network initialization.

As shown in Table~\ref{tab:extrap_class}, MLP mass predictions for heavy baryons are  slighlty less accurate than those for exotic candidates. On the other hand, GPR, XGBoost, and LightGBM predictions for heavy baryon masses are more accurate than exotic candidate masses. We find that the slightly larger heavy baryon error using MLP arises mainly from the ten excited singly heavy states (RMSE $331 \pm 63$~MeV), all of which are underestimated. On the other hand, the three doubly charmed baryons are reproduced more accurately ($168 \pm 126$~MeV). Although the quark contents of the singly heavy states appear in the training set, their masses lie well above the corresponding ground states, and the model underestimates these excitation energies.
For the singly heavy states, the tree-based models perform similarly, with RMSEs of $307 \pm 38$~MeV for XGBoost and $341 \pm 33$~MeV for LightGBM, whereas GPR gives $733 \pm 20$~MeV. For the doubly charmed baryons, however, the tree-based models reach about 1180~MeV and GPR $443 \pm 176$~MeV. The advantage of the MLP for heavy baryons therefore comes from the doubly charmed states, which require extrapolation to a second heavy quark. The results of this study indicate that good performance on the conventional hadron test set does not necessarily imply that ML models perform reliable extrapolation.

To understand the differences in extrapolation performance of ML models, we examine how each model learns from the training data and also extends this information beyond the training range. As it is known, tree ensembles with constant valued leaves produce piecewise constant predictions. Therefore, increasing a feature beyond its highest split threshold, when all other inputs are fixed, does not change the prediction of model. As explained in Sect.~\ref{subsec:ML}, since no training hadron contains more than one charm or bottom quark, every learned split on $n_c$ or $n_b$ separates zero from one. A hadron with two heavy quarks therefore follows the same path through each tree as one with a single heavy quark and receives the same prediction.  Most exotic candidates remain within the training range of each feature but combine values that do not occur together in training. Each tree assigns these inputs to a leaf fitted to conventional hadrons satisfying the same split conditions.  To sum up, these limitations may contribute to the large extrapolation errors of XGBoost and LightGBM.  

For GPR, extrapolation is controlled by the kernel and the prior mean. Using the Mat\'ern kernel, each prediction is effectively a weighted combination of the training masses, with weights given by correlations that decrease with distance from the training samples. As the correlations approach zero, the predicted standardized log-mass also approaches the zero prior mean, which corresponds to the training-set mean log-mass, or a mass of about 2.0~GeV. However, the fitted length scales are long relative to the distances between the extrapolated states and their nearest training hadrons. Therefore, these neighbours still carry substantial weight and the predictions move only partly towards this limit. For example, GPR predicts an average mass of 4.3~GeV for the doubly bottom baryons, between the 2.0~GeV limit and the 5.6--6.05~GeV masses of their nearest training hadrons, the singly bottom baryons, and far below the expected scale near 10~GeV.

Compared to other ML models considered in this study, the MLP learns a single global function of the input features rather than a local fit to the training data. Trends learned from conventional hadrons, such as the mass increase associated with each heavy quark, can therefore continue beyond the training range. This extension is, however, not constrained by physics: it depends on the training partition, initialization, and regularization, and saturation of the $\tanh$ activation may weaken the network's response far from the training data. Nonetheless, it is seen that MLP extrapolates most reliably for states close to the training distribution, such as the doubly charmed baryons, and becomes less accurate and less stable further away. Since the MLP gives the lowest extrapolation RMSEs, we focus on its predictions hereafter.

Figures~\ref{fig:spectrum_exotic} and~\ref{fig:spectrum_heavy} compare the MLP mass predictions with experimental values from the 2026 PDG~\cite{ParticleDataGroup:2026mpi} and the LHCb measurements of $\Xi_{cc}^+$ and $\Omega_{cc}^+$~\cite{LHCb_Xicc_plus_2026,LHCb_Omegacc_plus_2026}. In addition to experimental data, we include theoretical predictions for comparison from quark models~\cite{Gal:2022yqu, Lu:2020rog, Roberts:2007ni, Yoshida:2015tia, Ebert:2007nw, PhysRevD.90.094007, PhysRevD.97.094006, PhysRevD.110.056052}, a coupled-channel calculation~\cite{PhysRevD.108.074012}, QCD sum rules~\cite{Agaev2018, PhysRevD.78.094007}, and lattice QCD~\cite{PhysRevD.90.094507}. One should note that in Ref. \cite{PhysRevD.108.074012}, authors fitted overall short-range interaction strength to the experimental masses of \(P_{cs}(4338)\), \(P_{cs}(4455)\), and \(P_{cs}(4468)\). Therefore, the corresponding mass comparison involves an experimentally calibrated model.

As mentioned above, the MLP achieves the lowest extrapolation RMSE among the four ML models. For the exotic candidates, it reproduces the broad mass pattern well. Several predictions also lie close to the experimental values. We note that the MLP underestimates the masses of $Y(4360)$, $Y(4660)$, two $Z_b$ states, and $P_c(4312)^+$ by about $0.3$--$0.5$~GeV. Nonetheless, the results are comparable to the theoretical calculations and the model can make reasonable predictions for unseen hadrons.

As can be seen from Fig.~\ref{fig:spectrum_heavy}, the MLP also reproduces the broad mass pattern for heavy baryons. MLP predictions are in close agreement with the experimental data for $\Xi_b(6095)^0$ and $\Omega_b(6340)^-$. We find that the remaining heavy baryons are underestimated by 0.15--0.45~GeV, with the largest deviations, about 0.4~GeV, for $\Omega_b(6316)^-$, $\Xi_b(6327)^0$, and the two $\Omega_c$ states. Despite these deviations, the MLP gives the lowest heavy baryon RMSE among the four models considered in this study.

Given MLP's performance on the measured out-of-sample states, we also examine its predictions for nine unobserved doubly heavy baryons. In Fig.~\ref{fig:spectrum_nonobserved}, we compare our predictions with QCD sum rules~\cite{PhysRevD.78.094007}, lattice QCD~\cite{PhysRevD.90.094507}, and Karliner--Rosner quark-model estimates~\cite{PhysRevD.90.094007, PhysRevD.97.094006}, and CQM \cite{Roberts:2007ni}. In the doubly charmed sector, MLP predicts masses of approximately $3.6$--$3.7$~GeV for the $\Xi_{cc}^{*}$ states and $\Omega_{cc}^{*+}$. These values are relatively close to several theoretical estimates, although their central values are lower. In the doubly bottom sector, MLP predicts masses of approximately $9.3$--$9.4$~GeV, which are below the theoretical values.

\subsection{Explainability analysis}
\label{subsec:shap}
Our results show that MLP can predict hadron masses with reasonable accuracy in both the interpolation and extrapolation regions. However, prediction errors alone cannot explain how the model uses the input features to predict hadron masses. Therefore, in this section, we use Shapley values~\cite{Lundberg:2017} to examine the feature contributions for MLP.  This analysis reveals the individual contribution of each feature to the predicted hadron masses.  For an input $\mathbf{x}$, we express the prediction for the hadron mass as

\begin{equation}
m(\mathbf{x})=\phi_0+\sum_i\phi_i(\mathbf{x}),
\end{equation}

\noindent where $\phi_0$ is the mean network output over a background set of training hadrons, and $\phi_i$ is the interventional Shapley value of feature $i$. An interventional Shapley value measures a feature's contribution to a prediction by comparing model outputs when that feature takes its actual value versus values from the background data. Within this framework, positive SHAP values increase the predicted mass above the baseline, while negative values decrease it. 

\begin{figure}[th!]
    \centering    \includegraphics[width=\linewidth]{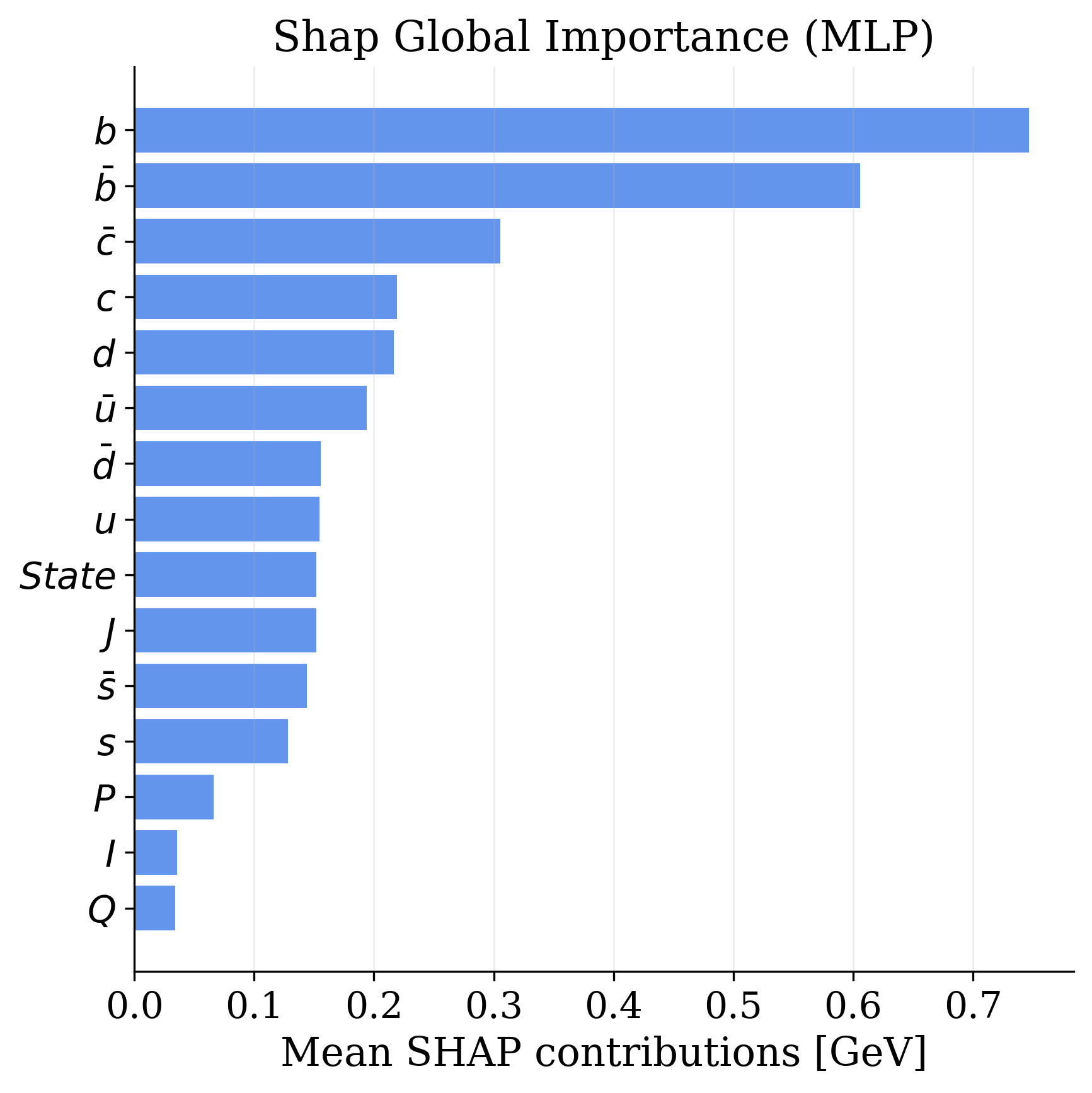}
    \caption{Global feature importance of the MLP, measured by the mean absolute exact Shapley value over 432 conventional hadrons. Values are given in GeV.} 
    \label{fig:shap_global}
\end{figure}

\begin{figure*}[th!]
    \centering
\includegraphics[width=\linewidth]{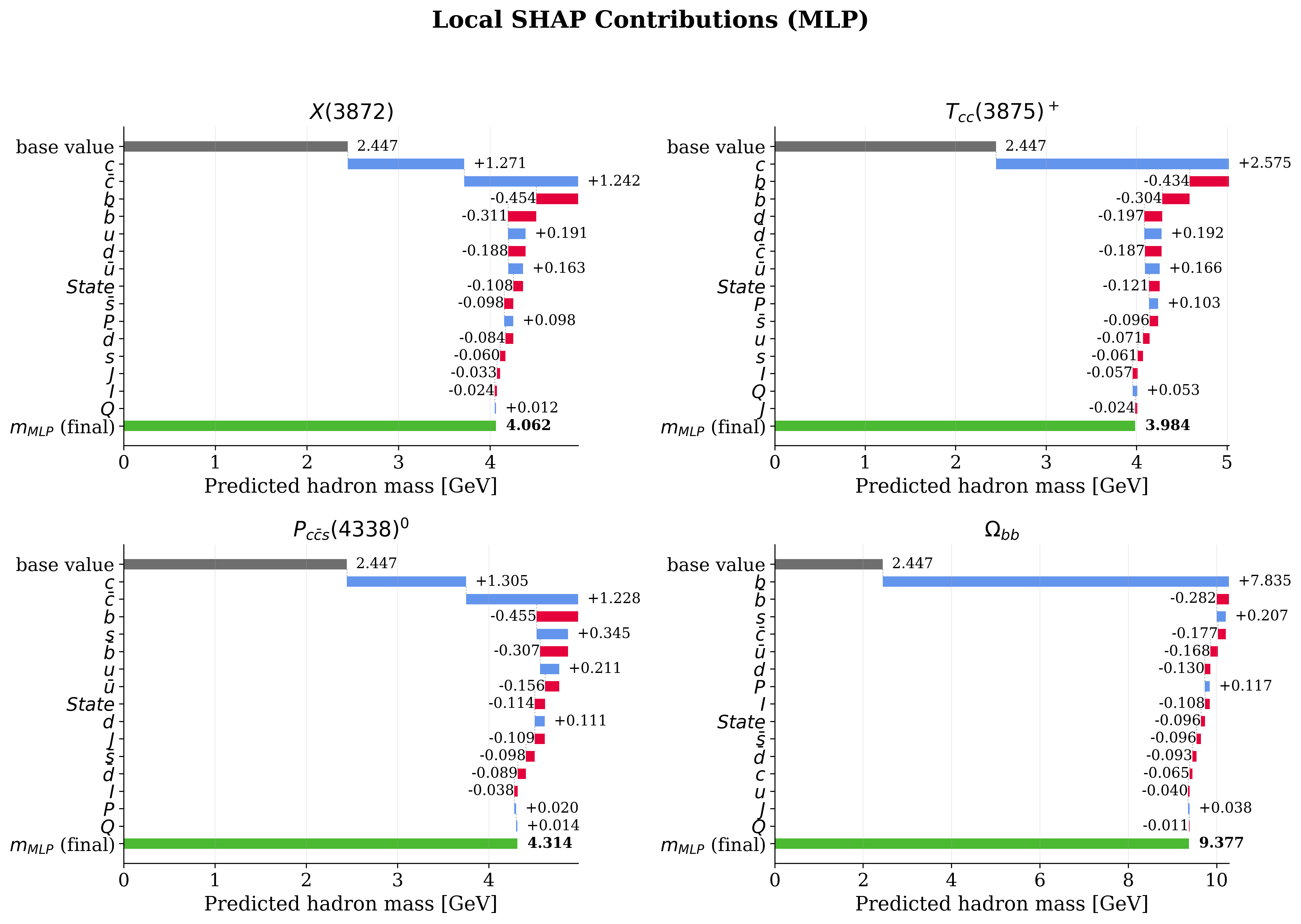}
    \caption{Local exact Shapley decompositions for the MLP for $X(3872)$, $T_{cc}(3875)^+$, $P_{c\bar c s}(4338)^0$, and $\Omega_{bb}$. Blue (red) bars increase (decrease) the prediction relative to the base value; the green bar is the final predicted mass. The SHAP values are computed for the reference split network and the final masses slightly differ from the ten-split averages.}
    \label{fig:shap_local_15}
\end{figure*}

In Fig.~\ref{fig:shap_global}, we show the mean absolute Shapley contributions $\langle|\phi_i|\rangle$ of the MLP inputs. We find that the bottom quark and bottom antiquark features contribute most, followed by the charm features. Light and strange quark features, the state, and $J$ make smaller contributions. The features $P$, $I$, and $Q$ contribute least. Our results indicate that the MLP relies mainly on heavy quark content to set the predicted mass scale.

We further examine the impact of these features on individual mass predictions. To this end, we calculate exact local SHAP values for four representative states outside the training data. To this end, we choose the exotic candidates $X(3872)$, $T_{cc}(3875)^+$ and $P_{c\bar{c}s}(4338)^0$ and the doubly bottom baryon $\Omega_{bb}$. Figure~\ref{fig:shap_local_15} shows the local exact Shapley decompositions for the MLP. 
We start from the mean model output over the background set of 30 training hadrons, which is shown by the gray bar at the top of each panel and gives 2.447 GeV. Then, each feature either increases (blue) or decreases (red) the mass prediction on top of this base value. The final predicted mass, which is shown by a green bar at the bottom of every panel, is the sum of the base value and feature contributions.

We find that heavy quark features dominate the MLP predictions. For $X(3872)$ and $P_{c\bar{c}s}(4338)^0$, the $c$ and $\bar c$ features make similar positive contributions. For $T_{cc}(3875)^+$, the mass prediction is also driven by the $c$, while the $\bar c$ contribution is comparatively small and negative. For $\Omega_{bb}$, most of the increase relative to the baseline is driven by the $b$ feature.

Features with zero input values for a selected hadron can also have nonzero SHAP contributions because SHAP measures contributions relative to the background data. We find that for the three charmed states, which contain no bottom quarks or antiquarks, the $b$ and $\bar b$ features consistently lower the predictions relative to the baseline.  We also find that light quark features produce smaller shifts and generally no more than a few hundred MeV each. For $P_{c\bar{c}s}(4338)^0$, the $s$ feature makes a positive contribution which is consistent with its assigned $udsc\bar c$ content. The quantum-number features $J$, $I$, $P$, and $Q$ also make comparatively small contributions to the three charmed-state predictions.

In conclusion, we show that MLP model mainly relies on heavy quark content to predict hadron masses. Their local contributions match the heavy quark compositions of the chosen out-of-sample states. This shows that the MLP model maintains this dependency throughout their extrapolation. These findings help us to explain the models' predictions without establishing a physical mechanism for hadron masses.

\section{Conclusion}
\label{sec:discuss}

In this study, we assess the predictive performance of four machine learning models---MLP, GPR, XGBoost, and LightGBM---for hadron masses, with particular attention to exotic candidates and heavy baryons. We use a dataset of $432$ conventional meson and baryon entries with experimentally measured masses and 15 features describing valence quark content and hadron quantum numbers. Then, we evaluate the predictive performance of models on held-out test sets and a separate extrapolation set of $39$ states. The extrapolation set contains exotic candidates and heavy baryons, none of which appear in the training or test sets.

We show that all four ML models learn a useful mapping from physically motivated descriptors to hadron masses and predict conventional hadron masses with reasonable accuracy within the interpolation regime. However, the performance of these ML models on the separate extrapolation set shows that this accuracy does not necessarily extend to unseen quark configurations.

We find that with just one hidden layer containing 16 neurons, the MLP achieves the lowest extrapolation RMSE among the four models in both the exotic and heavy baryon sectors. Across the 30 extrapolation states with experimental data the RMSE is obtained as $294.2 \pm 57.9$ MeV. That's about one third of the 780.9-877.7 MeV obtained with the other ML models. We also find that the advantage of MLP is largest for states with heavy quark content absent from training, such as the doubly charmed baryons. The main reason behind this is that tree-based models cannot distinguish one heavy quark from two, and GPR predictions move towards the training-set mean.  We also note that for the excited singly heavy baryons, whose quark contents appear in the training, the tree-based models reach a similar accuracy. For the nine unobserved doubly heavy baryons, it is shown that the MLP predictions in the doubly charmed sector lie slightly below theoretical estimates.  Finally, we use SHAP analysis and identified heavy quark content as the dominant contributor to MLP mass predictions, as expected given its role in setting the hadron mass scale.

These findings support the use of ML models as complementary tools for hadron spectroscopy. Future studies by adding more physics-informed features and theory-based constraints may improve predictions for hadrons and guide future experiments in the field. We will investigate these possibilities in future work.

\begin{acknowledgements}
E.Y. acknowledges support from the UK STFC under award no. ST/Y000358/1.
\end{acknowledgements}


\setlength{\bibsep}{-0.5pt}
\bibliographystyle{spphys-hep}
\bibliography{references}

\end{document}